\documentstyle[prl,aps,multicol,epsf]{revtex} 
\tighten
\begin{document}
\draft
\title{Negative Magnetoresistance of Granular Metals in a Strong
Magnetic Field} \author{I.S. Beloborodov$^{(1)}$ and
K.B. Efetov$^{(1,2)}$} \address{$^{(1)}$Theoretische Physik III,
Ruhr-Universit\"{a}t Bochum, 44780 Bochum, Germany\\
$^{(2)}$L.D. Landau Institute for Theoretical Physics, 117940 Moscow,
Russia \\ } \date{\today} \maketitle

\begin{abstract} 
The magnetoresistance of a granular superconductor in a strong
magnetic field destroying the gap in each grain is considered. It is
assumed that the tunneling between grains is sufficiently large such
that all conventional effects of localization can be neglected. A
non-trivial sensitivity to the magnetic field comes from
superconducting fluctuations leading to the formation of virtual
Cooper pairs and reducing the density of states. At low temperature,
the pairs do not contribute to the macroscopic transport but their
existence can drastically reduce the conductivity. Growing the
magnetic field one destroys the fluctuations, which improves the
metallic properties and leads to the negative magnetoresistance.
\end{abstract} 
 
 
\begin{multicols}{2} 
 
In a recent experiment \cite{Gerber97}, the resistance of a system of
$Al$ grains as a function of the magnetic field was studied. The
samples were quite homogeneous with a typical diameter of the grains
$120\pm 20\AA $ and the grains formed a $3$-dimensional
array. Destroying the superconductivity by the magnetic field the
authors could observe that in a sufficiently strong magnetic field the
system had a finite resistivity. The applied magnetic fields reached
$17T$, which was more than sufficient to destroy also the
superconducting gap in each grain. 

The dependence of the resistivity on the magnetic field observed in
Ref.  \cite{Gerber97} was not simple. At strong magnetic fields, the
resistivity {\it increased }when decreasing the magnetic field. Only
at sufficiently weak magnetic fields the resistivity decreased and
finally the samples displayed superconducting properties. A similar
behavior had been reported in a number of publications
\cite{Rutgers,Gantmakher96}. A negative magnetoresistance due to weak
localization effects is not unusual in disordered metals
\cite{lee}. However, the magnetoresistance of the granulated materials
is quite noticeable in magnetic fields exceeding $10T$, which is many
orders of magnitude higher than the typical values relevant for the
weak localization.
 
The aim of this Letter is to demonstrate that the magnetoresistance of
a granulated metal in a strong magnetic field and at low temperature
{\it must be} {\it negative due to superconducting} {\it
fluctuations.} We consider a system of granules coupled to each
other. The tunnelling amplitude between the grains is assumed to be
large, such that the system without electron-electron interactions
would be macroscopically a good metal and would not be sensitive to a
magnetic field. The superconducting gap in each granule is assumed to
be suppressed by the strong magnetic field. All the interesting
behavior considered below originates from the superconducting
fluctuations that lead to a suppression of the density of states (DOS)
but do not help to carry an electric current.
 
Theory of superconducting fluctuations near the transition into the
superconducting state has been developed long ago \cite
{Aslamazov68,Maki68,Abrahams} (for a review see
Ref. \cite{Varlamov}). Above the transition temperature $T_{c}$,
non-equilibrium Cooper pairs are formed and a new channel of charge
transfer opens \cite{Aslamazov68}. Another fluctuation contribution
comes from a coherent scattering of the electrons forming a Cooper
pair on impurities \cite{Maki68}. Both the fluctuation corrections
increase the conductivity and lead to a positive
magnetoresistance. Formation of the non-equilibrium Cooper pairs
results also in a fluctuational gap in the one-electron spectrum
\cite{Abrahams} but in conventional superconductors the first two
mechanisms are more important and the conductivity increases when
approaching the transition. A small decrease of the transverse
conductivity is possible in layered materials \cite{ioffe} in a
temperature interval not very close to the transition. It is relevant
to emphasize that the study of the fluctuations has been done near the
critical temperature $T_{c}$ in a zero or a weak magnetic field.
 
In granulated materials, the superconducting gap in each granule can
be destroyed at low temperature by a strong magnetic field. At
magnetic fields $H$ exceeding the critical field $H_{c}$ virtual
Cooper pairs can still be formed. However, as it will be shown below,
the influence of these pairs on the macroscopical transport is
drastically different from that near $T_{c}$.  The existence of the
virtual pairs leads to a reduction of the DOS but, in the limit
$T\rightarrow 0,$ these pairs cannot travel from one granule to
another. As a result, the conductivity $\sigma $ can be at $H>H_{c}$
considerably lower than conductivity $\sigma _{0}$ of the normal metal
without an electron-electron interaction. It approaches the value
$\sigma _{0}$ only in the limit $H\gg H_{c},$ when all the
superconducting fluctuations are completely suppressed by the magnetic
field.
 
For explicit calculations we assume that the superconducting pairing
inside the grains is mainly destroyed by the orbital mechanism. The
Zeeman splitting is not important provided the radius $R$ of the
grains is large enough, so that $R\gg \xi \left( \varepsilon _{0}\tau
\right) ^{-1}$, where $ \xi$ is the superconducting coherence length,
$\varepsilon _{0}$ is the Fermi energy and $\tau $ is the elastic mean
free time. This condition is well satisfied in grains with $R\sim
100\AA $ studied in Ref.  \cite{Gerber97}. This limit is opposite to
the one considered recently in Ref.\cite{Aleiner97} where the Zeeman
splitting was assumed to be the main mechanism of destruction of the
Cooper pairs.
 
We assume that electrons can hop from grain to grain and the tunneling
energy $t$ is in the interval
\begin{equation} 
\delta \ll t\ll \Delta _{0}  \label{a1} 
\end{equation} 
where $\delta =\left( \nu V\right) ^{-1}$ is the mean level spacing in
a single granule, $\nu =mp_{0}/2\pi ^{2}$ is the DOS of the metal in
the absence of interactions, $V$ is the volume of the granule, and
$\Delta _{0}$ is the BCS gap at $T=0$ in the absence of a magnetic
field. The left inequality in Eq. (\ref{a1}) guarantees that the
system is macroscopically a good metal \cite{Efetov} and localization
effects can be neglected.  Moreover, charging effects are also not
important in this limit. In the limit $R\ll \xi $
considered below the superconducting fluctuations in a
single grain are effectively zero-dimensional ($0D$).
 
The Hamiltonian $\hat{H}$ of the system can be written as  
\begin{equation} 
\hat{H}=\hat{H}_{0}+\hat{H_{T}},  \label{fulH} 
\end{equation} 
where $\hat{H}_{0}$ is a conventional Hamiltonian with an
electron-phonon interaction in the presence of a strong magnetic
field. The term $\hat{H}_{T} $ in Eq.(\ref{fulH}) describes the
tunneling from grain to grain and has the form (see
e.g. Ref.\cite{Kulik})
\begin{equation} 
\hat{H}_{T}=\sum\limits_{i,j,p,q}t_{ijpq}a_{ip}^{\dag }a_{jq}+h.c. 
\label{tun} 
\end{equation} 
where $a_{ip}^{\dag }\left( a_{ip}\right) $ are creation (annihilation)
operators for an electron the grain $i$ and state $p$.
 
Correspondingly, the a.c. current density ${\bf j}\left( t\right) $ is 
written using linear response formulae as  
\begin{equation}  
{\bf j}(t) =i \int\limits_{-\infty }^{t}\langle \lbrack {\bf  
\hat{\jmath}}(0,t),{\bf \hat{\jmath}}(0,t^{\prime })]\rangle {\bf A} 
(t^{\prime })dt^{\prime },  
\end{equation}
where ${\bf \hat{\jmath}}$ is the tunneling current operator and the
angle brackets stand for averaging over both quantum states and
impurities in the grains. In principle, the grains can be clean and
the electrons can scatter mainly on the surface of the
grains. However, provided the shape of the grains corresponds to a
classically chaotic motion of the electrons, the clean limit should be
described in the $0D$ case by the same formulae.
 
We carry out the calculation of the conductivity making expansion both
in fluctuation modes and in the tunneling term $H_{T}$.  The right
part of Eq. (\ref{a1}) can provide a small parameter of the
expansion in the tunneling.  As in conventional superconductors, we
can single out contributions due to corrections to the DOS and
Aslamazov-Larkin (AL) $\sigma _{AL}$ and Maki-Thompson (MT) $\sigma
_{MT}$ corrections. Diagrams describing these contributions are
represented in Fig. 1. As a result, the total conductivity $\sigma $
can be written as
\begin{equation} 
\sigma =\sigma _{DOS}+\sigma _{AL}+\sigma _{MT}  \label{a2} 
\end{equation} 
where $\sigma _{DOS}$ is given by equation
\begin{equation} 
\sigma _{DOS}=\sigma _{0}\left( 4T\right) ^{-1}\int\limits_{-\infty
}^{+\infty }[\nu (\varepsilon )/\nu _{0}]^{2}\cosh
^{-2}(\frac{\varepsilon }{ 2T})d\varepsilon , \label{a3}
\end{equation} 
$\sigma _{0}=4\pi e^{2}d^{-1}\left( t/\delta \right) ^{2}$ is the
classical conductivity of the granular metal and $\nu \left(
\varepsilon \right) $ is the density of states modified by the
superconducting fluctuations and averaged over the impurities.
 
The main correction $\delta \nu \left( \varepsilon \right) $ to the
DOS of the non-interacting electrons $\nu _{0}$ is described by the
diagram in the Fig. 1a, while the terms $\ \sigma _{MT}$ and $\sigma
_{AL}$ are given by Figs. 1b and 1c, respectively. The calculation of
the diagrams can bee performed for the Matsubara frequencies
$\varepsilon _{n}=\pi T(2n+1)$ using temperature Green functions. At
the end one should, as usual \cite{AGD}, make the analytical
continuation $i\varepsilon _{n}\rightarrow \varepsilon $ . The
magnetic field can be considered in the quasiclassical approximation,
which results in additional phases in Green functions.
 
The diagrams in Fig.1 contain essentially the averaged one-particle
Green functions, the impurity vertices $C$ and the propagator of the
superconducting fluctuations $K$. The functions $C$ and $K$ depend on
the coordinates and time slower than the averaged one-particle Green
functions.  As a result, the magnetic field affects only the vertex
$C$ and the propagator $K$, whereas the phases of the Green functions
drawn in Fig.1 outside these blocks cancel. So, reading the diagrams
in Fig.1 one should replace the solid lines by the functions
\begin{equation} 
G^{\left( 0\right) }\left( i\varepsilon _{n},{\bf p}\right) =\left( 
i\varepsilon _{n}-\xi \left( {\bf p}\right) +i\left( 2\tau \right) 
^{-1}sgn\varepsilon _{n}\right) ^{-1}  \label{a5} 
\end{equation} 
 
The impurity vertex entering these diagrams is equal to $\left( 2\pi
\nu _{0}\tau \right) ^{-1}C\left( i\varepsilon _{n},i\Omega
_{k}-i\varepsilon _{n}\right) $, where $C$ is the so-called
Cooperon. It obeys the following equation
\begin{eqnarray} 
&&\left( D_{0}\left( -i{\bf \nabla +}\left( 2e/c\right) {\bf A}\right) 
^{2}+|2\varepsilon _{n}-\Omega _{k}|\right) C\left( {\bf r,r}^{\prime 
}\right)  \label{a6} \\ 
&=&2\pi \nu _{0}\delta \left( {\bf r-r}^{\prime }\right)  \nonumber 
\end{eqnarray} 
where $D_{0}=v_{0}^{2}\tau /3$ is the classical diffusion
coefficient. The vector-potential ${\bf A}\left( {\bf r}\right) $
should be chosen in the London gauge. If the shape of the grain is
close to spherical, the vector-potential is expressed through the
magnetic field ${\bf H}$ as $ {\bf A}\left( {\bf r}\right) =[{\bf
H\times r]/}2$.
 
All relevant energies in the problem are assumed to be much smaller
than the energy of the first harmonics $E_{c}=D_{0}\pi ^{2}/R^{2}$
playing the role of the Thouless energy of a single grain and this
allows to keep only the zero harmonics in the spectral expansion. One
can find the eigenvalue ${\cal E}_{0}\left( H\right) $ of this
harmonics using the first order of the standard perturbation theory
\begin{equation} 
{\cal E}_{0}\left( H\right) =\left( 2e/c\right) ^{2}D_{0}<{\bf
A}^{2}>_{0}
\label{a8} 
\end{equation} 
where $<...>_{0}$ stands for the averaging over the volume of the
grain. For a grain of a nearly spherical form one obtains
\begin{equation} 
{\cal E}_{0}\left( H\right) =\frac{2}{5}\left( \frac{eHR}{c}\right)
^{2}D_{0}=\frac{2}{5}\left( \frac{\phi }{\pi \phi _{0}}\right)
^{2}E_{c}
\label{a9} 
\end{equation} 
where $\phi _{0}=\pi c/e$ is the flux quantum and $\phi $ is the
magnetic flux through the granule.
 
Within the zero-harmonics approximation, the function $C$ does not
depend on coordinates and equals
\begin{equation} 
C\left( i\varepsilon _{n},i\Omega _{k}-i\varepsilon _{n}\right) =2\pi
\nu _{0}\left( |2\varepsilon _{n}-i\Omega _{k}|+{\cal E}_{0}\left(
H\right) \right) ^{-1} \label{a10}
\end{equation} 
 
The propagator of the superconducting fluctuations is calculated
summing, as usual, the superconducting ladder containing the products
$ G_{i\varepsilon _{n}}G_{i\Omega _{k}-i\varepsilon _{n}}.$ Neglecting
weak localization corrections one reduces the averaging of the
propagator $K$ over the impurities to the averaging of these products
that give finally $C$. The tunneling can be neglected here provided
the right part of Eq. (\ref{a1}) is fulfilled. However, the tunneling
from grain to grain should be taken into account when calculating
$K$. This can be done making expansion in $H_{T}$,
Eq. (\ref{tun}). Each new vertex arising in this expansion should be
also dressed by impurity lines.
 
Although the final result can be written for arbitrary $T$ and $H$,
let us concentrate on the most interesting case $T\ll T_{c}$,
$H>H_{c}$. Assuming for simplicity that the granules are packed into a
cubic lattice and using the momentum representation with respect to
coordinates of the grains we obtain in this limit
 
\[ 
K(i\Omega _{k},{\bf q})=-\nu _{0}^{-1}\left( \ln \left( \frac{{\cal
E}_{0}(H)}{\Delta _{0}}\right) +\frac{|\Omega _{k}|}{{\cal
E}_{0}(H)}+\eta ({\bf q} )\right) ^{-1},
\] 
 
\begin{equation} 
\eta ({\bf q})\equiv \left( 4/3\pi \right) \sum_{i=1}^{3}J\left(
\delta / {\cal E}_{0}(H)\right) \left( 1-\cos q_{i}d\right) ,
\label{eta}
\end{equation} 
where $J=\left( \pi ^{2}/4\right) \left( t/\delta \right) ^{2}$, ${\bf
q}$ is the quasi-momentum, $d=2R$, and $|\Omega _{k}|\ll {\cal
E}_{0}(H)$. The pole of the propagator $K$, Eq. (\ref{eta}), at ${\bf
q}=0$, $\Omega _{k}=0$ determines the field $H_{c}$, at which the BCS
gap disappears
\begin{equation} 
{\cal E}_{0}(H_{c})=\Delta _{0}  \label{a11} 
\end{equation} 
 
The result for $H_{c}$, Eqs. (\ref{a9}, \ref{a11}), agrees with the
one obtained long ago by another method \cite{Larkin}. The term $\eta
({\bf q})$ in Eq.(\ref{eta}) is very important if $H$ is close to
$H_{c}$. For $ \varepsilon _{n}>0$ the contribution of the diagram
Fig. 1a, is:
\begin{equation} 
\delta \nu (i\varepsilon _{n})=\frac{2iT}{\nu _{0}}\sum\limits_{\Omega
_{k}<\varepsilon _{n}}\int K(i\Omega _{k},{\bf q})C^{2}\left(
i\varepsilon _{n},i\Omega _{k}-i\varepsilon _{n}\right)
\frac{d^{3}q}{\left( 2\pi \right) ^{5}} \label{fulnu}
\end{equation} 
After calculation of the sum over $\Omega _{k}$ in Eq. (\ref{fulnu}),
one should make the analytical continuation $i\varepsilon
_{n}\rightarrow \varepsilon $. At low temperatures it is sufficient to
find $\delta \nu \equiv \delta \nu \left( 0\right) $.
 
Remarkably, Eqs. (\ref{eta}-\ref{fulnu}) do not contain explicitly the
mean free time $\tau $. This is a consequence of the zero-harmonics
approximation, which is equivalent to using the random matrix theory
(RMT) \cite{Efetov}. (The parameter $\tau $ enters only Eq. (\ref{a9})
giving the standard combination ${\cal E}_{0}\left( H\right) $
describing in RMT the crossover from the orthogonal to the unitary
ensemble). This justifies the claim that the results can be used also
for clean grains with a shape providing a chaotic electron motion.
 
Using Eqs. (\ref{a3}, \ref{eta}-\ref{fulnu}) one can easily obtain an
explicit expression for $\sigma _{DOS}$ for $H-H_{c}\ll H_{c}$. In
this limit, one expands the logarithm in the denominator of
Eq.(\ref{eta}) and neglects the dependence of $C$ on $\varepsilon
_{n}$ and $\Omega _{k}$ because the main contribution in the sum over
$\Omega _{k}$ comes from $ \Omega _{k}\sim {\cal E}_{0}(H)-{\cal
E}_{0}(H_{c})\ll \Delta _{0}$. The result for $\delta \sigma
_{DOS}=\sigma _{0}-\sigma _{DOS}$ can be written as
\begin{equation} 
\frac{\delta \sigma _{DOS}}{\sigma _{0}}=-\frac{2\delta }{\Delta
_{0}} \left\{
\begin{array}{lr} 
-\pi ^{-1}<\ln \tilde{\eta}\left( {\bf q}\right) >_{q}, & T/\Delta
_{0}\ll \tilde{\eta} \\ \frac{2T}{\Delta _{0}}<\tilde{\eta}^{-1}({\bf
q})>_{q}, & \tilde{\eta}\ll T/\Delta _{0}\ll 1
\end{array} 
\right. ,  \label{Hc} 
\end{equation} 
\[ 
\tilde{\eta}({\bf q})=\eta ({\bf q})+2h,\quad <...>_{q}\equiv V\int \left(
...\right) d{\bf q/}\left( 2\pi \right) ^{3}
\] 
where $h=\left( H-H_{c}\right) /H_{c}$. We see that the correction to
the conductivity is negative and its absolute value decreases when the
magnetic field increases. The correction reaches its maximum at
$H\rightarrow H_{c}$. At zero temperature, the maximum value of
$\delta \sigma _{DOS}/\sigma _{0}$ is of order $\delta /\Delta
_{0}$. As temperature grows, the correction can become larger and
reach for $T\sim \Delta _{0}$ the order of magnitude of $J^{-1}$.
Both the values are smaller than unity because we work in the region
specified by Eq. (\ref{a1}) and this justifies the diagrammatic
expansion we use. The correction can become comparable with $\sigma
_{0}$ when $J\sim 1$, which would mean that we are not far from the
metal-insulator transition.  For such values of $J$ one can use
Eq. (\ref{Hc}) only for rough estimates.  Apparently, the parameters
of the samples of Ref. \cite{Gerber97} correspond to the region $J\sim
1$, $\delta / \Delta_{0} \sim 1$.
 
In the opposite limit $H\gg H_{c}$ the correction to $\sigma _{0}$ can
still be noticeable. In this case we can neglect the dependence of the
superconducting propagator, Eq.(\ref{eta}), on $\Omega _{k}$ and on
the tunneling term (because now $\ln \left( {\cal E}_{0}(H)/{\cal E}%
_{0}(H_{c})\right) \gg 1$) and finally obtain
\begin{equation} 
\delta \sigma _{DOS}/\sigma _{0}=-(1/3)\left( \delta /{\cal E}
_{0}(H)\right) \ln ^{-1}\left( {\cal E}_{0}(H)/\Delta _{0}\right) ,
\label{highfild} 
\end{equation} 
which means that in the region $H\gg H_{c}$ the correction to the
conductivity decays essentially as $\delta \sigma _{DOS}\sim H^{-2}$.
 
Let us emphasize that the correction to the conductivity coming from
the DOS remains finite in the limit $T\rightarrow 0$, thus indicating
the existence of the virtual Cooper pairs even at $T=0$.
\begin{figure} \narrowtext {
\epsfysize =3.5cm
\centerline{\epsfbox{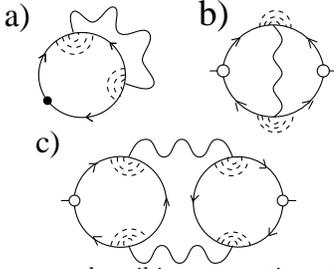}}}
\caption{Diagrams describing corrections to conductivity due to
superconducting fluctuations. The wavy lines denote the propagator of
the fluctuations, dashed lines describe
impurity scattering.}
\label{fig1} 
\end{figure} 
 
In order to calculate the entire conductivity, Eq. (\ref{a2}), we 
must investigate the AL and MT contributions (Figs. 1c and 1b). In 
conventional superconductors near $T_{c}$, these contributions are most 
important leading to an increase of the conductivity. 
 
In the granular materials the situation is much more interesting. It
turns out that both the AL and MT contributions {\it vanish} in the
limit $T\rightarrow 0$ at all $H>H_{c}$ and thus, the correction to
the conductivity comes from the DOS only. Leaving the presentation of
the details for a future publication \cite{prepar} we list here only
the final results.
 
The AL correction to the conductivity $\sigma _{AL}$ corresponds to
the tunneling of the virtual Cooper pairs and must be proportional to
$t^{4}$ in contrast to the one-electron tunneling determining $\sigma
_{DOS}$. All calculations are done in the same approximation as when
calculating $\sigma _{DOS}$. In the limit $T\ll T_{c}$, $H-H_{c}\ll
H_{c}$, the $\ $result for $ \sigma _{AL}$ can be written as
\begin{equation} 
\frac{\sigma _{AL}}{\sigma _{0}}=\frac{4\pi ^{2}t^{2}T}{9\Delta
_{0}^{3}} \sum\limits_{i=1}^{3}<A\left( {\bf q}\right)
[{\tilde{\eta}({\bf q})]}^{-3}{ \sin }^{2}q_{i}d>_{q} \label{fulAL}
\end{equation} 
where $A\left( {\bf q}\right) =4\pi T\left( 3\Delta
_{0}\tilde{\eta}({\bf q} )\right) ^{-1}$ for $T\ll \Delta
_{0}\tilde{\eta}$ and $A\left( {\bf q} \right) =1$ for $T\gg \Delta
_{0}\tilde{\eta}$ .
 
As usual, the MT diagrams have both regular and anomalous part. For
the problem considered, they are of the same order of magnitude but
have opposite signs. Moreover, at $T=0$ they cancel each other. The
final result for the MT contribution takes the form
\begin{equation} 
\frac{\sigma _{MT}}{\sigma _{0}}=\frac{8{\pi }^{2}T^{2}\delta
}{9\Delta _{0}^{3}}\sum\limits_{i=1}^{3}<B\left( {\bf q}\right) \cos
q_{i}d>_{q}
\label{fulMT} 
\end{equation} 
where $B\left( {\bf q}\right) =$ $-\pi ^{-1}\ln \tilde{\eta}({\bf q})$
for $ T\ll \Delta _{0}\tilde{\eta}$ and $B\left( {\bf q}\right)
=2T\left( \Delta _{0}\tilde{\eta}({\bf q})\right) ^{-1}$ for $T\gg
\Delta _{0}\tilde{\eta}$.
 
The temperature and magnetic field dependence of $\sigma _{AL}$ and
$\sigma _{MT}$ is rather complicated but they are definitely
positive. The competition between these corrections and $\sigma
_{DOS}$ determines the sign of the magnetoresistance. We see from
Eqs. (\ref{fulAL}, \ref{fulMT}) that both the AL and MT contributions
are proportional at low temperatures to $T^{2}$. Therefore the $\sigma
_{DOS}$ in this limit is larger and the magnetoresistance is negative
for all $H_{c}$. In contrast, at $T\sim T_{c}$ the AL and MT
corrections can become close to $H_{c}$ larger than $\sigma _{DOS}$
resulting in a positive magnetoresistance in this region. Far from $
H_{c}$ the magnetoresistance is negative again.
 
The present study cannot help to answer the question whether there is a 
phase transition at $H_{c}$ or not because the perturbation theory does not 
work very close to $H_{c}$. Investigation of this problem can bring a new 
understanding of the granular materials. 
 
In conclusion, we have shown that superconducting fluctuations in granular 
metals at low temperature and high magnetic field lead to a suppression of 
the density of states. The well known Aslamazov-Larkin and Maki-Thompson 
corrections vanish in the limit $T\rightarrow 0$ and all this leads to the 
reduction of the conductivity.  
 
The authors thank I. Aleiner, B. Altshuler and F. Hekking for helpful
discussions in the course of the work. A support of the
Graduiertenkolleg 384 and the Sonderforschungsbereich 237 is
greatly appriciated.

\end{multicols}
 
\end{document}